\documentclass[floatfix, onecolumn, pre, footinbib]{revtex4}

\usepackage{graphicx}        
\usepackage{psfrag}

\newcommand{\avg}[1]{\left\langle #1\right\rangle}

\begin{document}

\title{Traffic of molecular motors: from theory to
  experiments\footnote{Based on the oral contribution given at the
    Traffic and Granular Flow meeting 2007}} \author{Paolo Pierobon}


\affiliation{Institut Curie, CNRS UMR 168, 11 Rue P. et M.  Curie,
  F-75231 Cedex 05, France \footnote{Email adress: \tt{pierobon@curie.fr}}}

%
%
\maketitle

Intracellular transport along microtubules or actin filaments, powered
by molecular motors such as kinesins, dyneins or myosins, has been
recently modeled using one-dimensional driven lattice gases. We
discuss some generalizations of these models, that include extended
particles and defects. We investigate the feasibility of single
molecule experiments aiming to measure the average motor density and
to locate the position of traffic jams by mean of a tracer particle.
Finally, we comment on preliminary single molecule experiments
performed in living cells.

\section{Introduction}

Living cell is a highly organised structure that constantly needs to
move its constituent parts from one place to another. It is
therefore provided with a complex and accurate distribution system:
the \emph{cytoskeleton}, the network of biopolymers (actin,
microtubules and intermediate filaments) that gives the cell its
structural and mechanical features, functions as road system for the
transport of organelles and vesicles; the motion of these objects is
entrusted to \emph{motor proteins} moving along these filaments. These
motors are enzymes that convert the energy obtained by hydrolysis of
an ATP (adenosin-triphosphate) molecule into a work (displacement of
their cargo).  Myosin V (on actin filament) and kinesin or dynein (on
microtubules) are well known examples of these so called
\emph{processive} motors \cite{b:alberts-etal:02,b:howard:01}.

It has been observed that these motors can act cooperatively or
interact with each other giving rise to collective phenomena.  In
particular, in some phase of the cell cycle, the motors can be
expressed in high concentrations: it seems therefore natural to
investigate analogies and differences with the traffic observed in a
city. Experimental techniques as single molecule and fluorescence
imaging have just started giving some hints on the complex behaviour
of this system, but are still far from giving quantitative description
of traffic situations.  While waiting for experimental data some
theoretical models have been developed to physically describe
intracellular transport.

\section{Driven lattice gases: models for intracellular transport}
\label{sec:pff}
A simple model to capture the behavior of many motors on a filaments
needs to include three fundamental features: (i) the motors move in a
step-like fashion on a one dimensional tracks and bind specifically to
the monomer constituting the cytoskeletal filaments; (ii) cytoskeleton
filaments present a specific polarization: the chemical properties of
the track guarantee that the motion is always directed towards one of
the two ends of the filament; (iii) the particles move according to a
stochastic rule, i.e. they move upon a chemical reaction, the
hydrolysis of ATP, which occurs randomly with a typical rate.

The Total Asymmetric Simple Exclusion Process (TASEP) is a stochastic
process first introduced to describe the motion of ribosomes on mRNA
substrate \cite{macdonald-gibbs-pipkin:68} and encodes all these
features. It rapidly became a paradigm of non-equilibrium statistical
mechanics and one of the few example of exactly solvable systems
\cite{derrida-domany-mukamel:92,schuetz-domany:93}.  In this model
each particle occupies a site on a one dimensional lattice and
advances stochastically and in one direction.  The most obvious
observable is the average density profile of particles along the
lattice.

The system with open boundaries where particles enter the lattice with
rate $\alpha$ at one end and leave with rate $\beta$ at the other,
shows a non trivial phase diagram where three distinct non-equilibrium
steady states appear: a \emph{low density} phase controlled by the
left boundary, a \emph{high density} phase controlled by the right
boundary and a \emph{maximal current} phase, independent of the
boundaries.

\begin{figure}[htbp]
\centering
  \psfrag{alpha}{$\alpha$}
    \psfrag{beta}{$\beta$}
    \psfrag{a}{$a$}
    \psfrag{i=1}[c]{$i=1$}
    \psfrag{i=N}[c]{$i=N$}
    \psfrag{tau=1}[l]{$\tau=1$}
    \psfrag{omega_A}[][c]{$\omega_A$}
    \psfrag{omega_D}[][c]{$\omega_D$}
    \psfrag{a}{(a)}
    \psfrag{b}{(b)}

\includegraphics[width=\textwidth]{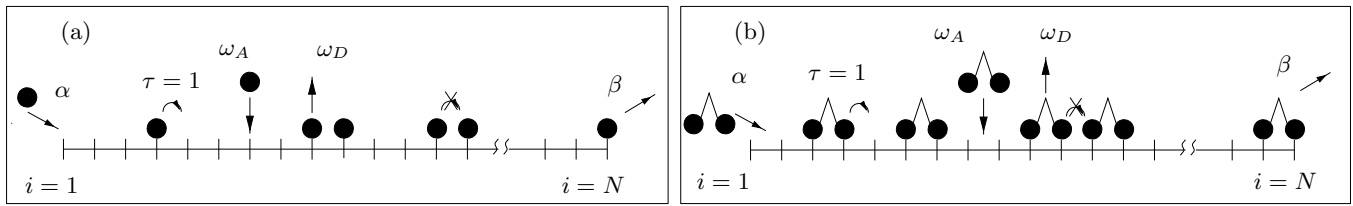}
\caption{Schematic representation of the TASEP with Langmuir kinetics in the
case of (a) monomers and (b) dimers. The allowed moves are: forward
jump (with rate $\tau=1$), entrance at the left boundary (with rate
$\alpha$), exit at the right boundary (with rate $\beta$), attachment
(with rate $\omega_A$), and detachment (with rate $\omega_D$) in the
bulk.}
\label{fig:model}       
\end{figure}

In a first attempt to construct a minimal model for molecular
intracellular transport, one needs to add to the TASEP the fact that
the tracks are embedded in the cytosol with a reservoir of motors in
solution.  This allows the motors to attach to or detach from the
track. This led to the construction of the TASEP with Langmuir (i.e.
attachment/detachment) kinetics or TASEP/LK model
\cite{parmeggiani-franosch-frey:03,parmeggiani-franosch-frey:04}
depicted in Fig.~\ref{fig:model}a.  According to this model, in
addition to the TASEP properties, particles enter (leave) the system
with rate $\omega_A$ ($\omega_D$) also in the bulk. All over the
lattice they obey exclusion: two particles cannot occupy the same
site.

According to the rules described above, the rate equation in the
average density at site $i=2\dots N$, $\avg{n_i}$ can be written as:
\begin{eqnarray}
  \label{eq:pff0}
  \partial_t \avg{n_i}=\avg{n_{i-1}\left(1-n_i\right)}-\avg{n_{i}\left(1-n_{i+1}\right)}+\left[\omega_A(1-\avg{n_i})-\omega_D\avg{n_i}\right]\, .
\end{eqnarray}
The first two brackets describe the average current and the second one
the on-off kinetics (source and sink terms). At the boundaries this
equation reads:
\begin{eqnarray}
 \partial_x
 \avg{n_1}&=&\alpha\left(1-\avg{n_i}\right)-\avg{n_{1}\left(1-n_{2}\right)}\, ,\\
 \partial_x
 \avg{n_N}&=&\avg{n_{N-1}\left(1-n_{N}\right)}-\beta\avg{n_N}\, .
\end{eqnarray}

\begin{figure}[htbp]
\centering

    \psfrag{r}[][][1][-90]{$\rho$}
    \psfrag{j}[][][1][-90]{$j$}
    \psfrag{x}{$x$}
    \psfrag{(a)}{(a)}
    \psfrag{(b)}{(b)}
    \psfrag{(c)}{(c)}    
    \psfrag{beta}{$\beta$}
    \psfrag{alpha}{$\alpha$}
    \psfrag{a*}[l]{$\alpha^*=1/2$}
    \psfrag{b*}[c]{$\beta^*=1/2$}
    \psfrag{LD}[][][0.8]{LD}
    \psfrag{HD}[][][0.8]{HD}
    \psfrag{LD/HD}[][][0.8]{LD/HD}
    \psfrag{MC/HD}[][][0.8]{MC/HD}
    \psfrag{LD/MC}[][][0.8]{LD/MC}
    \psfrag{MC}[][][0.8]{MC}
    \psfrag{LD/MC/HD}[][][0.8]{LD/MC/HD}
    \psfrag{X0}{$x_{w}=0$}
    \psfrag{X1}{$x_{w}=1$}
    \psfrag{0}[][bl]{0}
    \psfrag{1}[][bl]{1}
    \psfrag{ 0}[][bl]{0}
    \psfrag{ 1}[][bl]{1}
    \psfrag{ 0.2}[][bl]{0.2}
    \psfrag{xw}{$x_{w}$}
    \psfrag{ra}{$\rho_\alpha$}
    \psfrag{rb}{$\rho_\beta$}
    \psfrag{ja}{$j_\alpha$}
    \psfrag{jb}{$j_\beta$}    
\includegraphics[width=\textwidth]{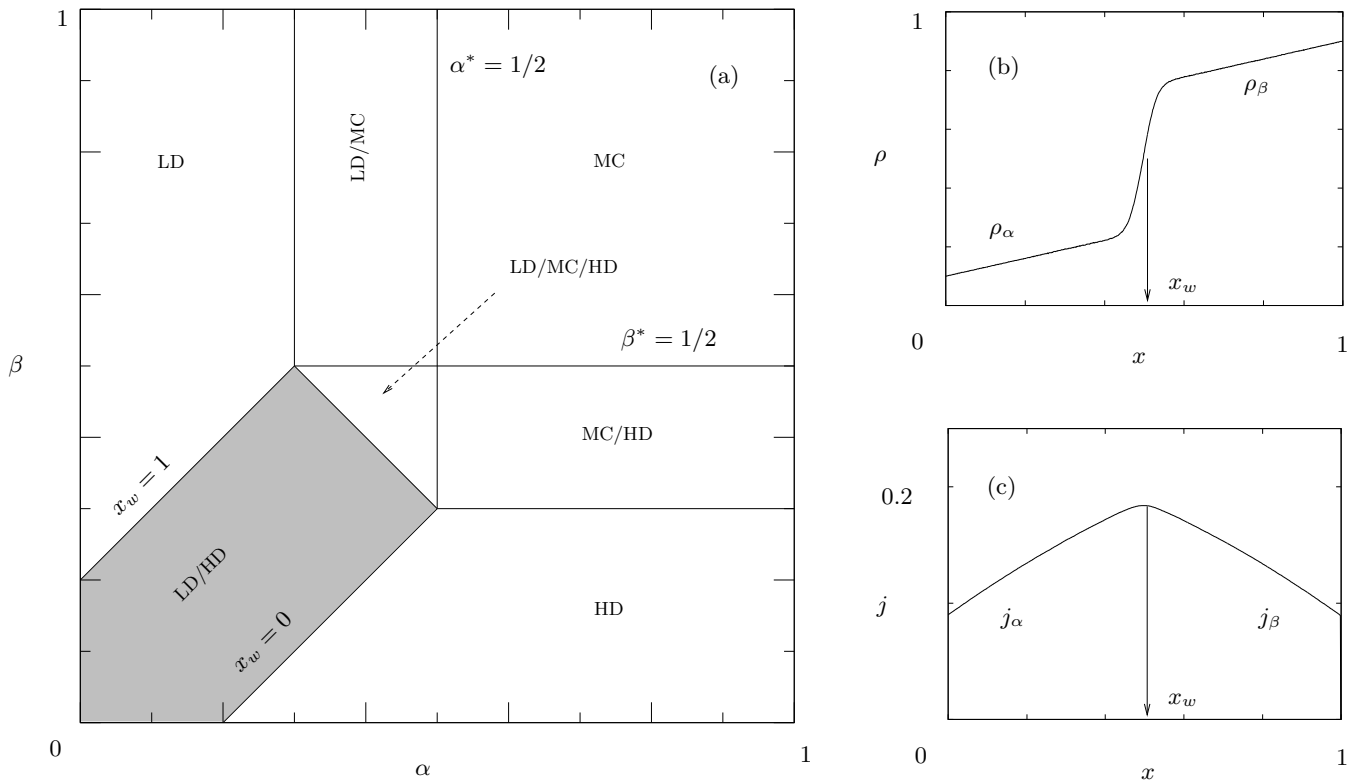}
\caption{  (a) Phase diagram of TASEP/LK for $\omega_A=\omega_D$.
  One recognizes seven phases: in addition to the TASEP low density
  (LD), high density (HD) and maximal current (MC) phases; there are
  four more coexistence phases, namely the LD/HD, LD/MC, MC/HD and
  LD/MC/HD phases.  The shaded region highlights the LD/HD coexistence
  where a localized domain wall appears. (b) Typical density profile
  in the LD/HD phase and (c) the corresponding current profile. At the
  matching point $x_{w}$ between the left ($j_\alpha$) and right
  ($j_\beta$) currents a domain wall develops and connects the left
  ($\rho_\alpha$) and right ($\rho_\beta$) density profile.}
\label{fig:taseplk}       
\end{figure}

Equation (\ref{eq:pff0}) shows a non-closed hierarchy in the
correlation functions (i.e.  $\avg{n_i}$ depends on $\avg{n_in_j}$ and
so on): this suggests the use of the mean field approximation
$\avg{n_i n_{i+1}}\approx\avg{n_i}\avg{n_{i+1}}$. In the stationary
state, the leading term of the continuum limit ($N\to\infty$, 
$\avg{n_i}\to\rho(x)$) of Eq.~(\ref{eq:pff0}) reads:
\begin{eqnarray}
  \label{eq:pff1}
  -\partial_x\left[\rho(1-\rho)\right]+\left[\Omega_A(1-\rho)-\Omega_D\rho\right]=0\,
  ,
\end{eqnarray}
supplemented by two boundary conditions: $\rho(0)=\alpha$ and
$\rho(1)=1-\beta$ \cite{parmeggiani-franosch-frey:03}. When the
solution of Eq.~(\ref{eq:pff1}) cannot be matched continuously with
the left and right boundaries, the density profile displays a
localized discontinuity (or \emph{shock}) in the bulk
(Figs.~\ref{fig:taseplk}b-c).  This translates into the emergence of
mixed phases that were not present in the simple TASEP.  For some sets
of parameters the phase diagram can exhibit up to 7 kinds of
coexistence (see Fig.\ref{fig:taseplk}a).

A key point in the study of TASEP/LK is the introduction of a
\emph{mesoscopic limit} where local adsorption-desorption rates
$\omega_{A,D}$ have been rescaled in the limit of large but finite
systems $\omega_{A,D}=\Omega_{A,D}/N$, such that the macroscopic rates
are comparable to the injection-extraction rates at the boundaries
\cite{parmeggiani-franosch-frey:03}. Far from being only a simple
mathematical trick that allows the competition of the directed motion
with the on-off kinetics, this limit captures the fact that the motors
explores a significant fraction of the track before detaching.  This
is precisely the limit of highly processive motors that biologically
motivated our studies.  Surprisingly only in the mesoscopic limit the
density profiles show the shock.

\section{Extending the model: dimers and defects}
\subsection{Dimers}
Many processive molecular motors (kinesins, dyneins and myosin V) are
composed of two heads that bind specifically each to a subunit of the
molecular track. A natural extension of the previous model towards
a more realistic one consists in introducing non-pointlike particles in
the system such as \emph{dimers} (see
Ref. \cite{pierobon-frey-franosch:06} and Fig.~\ref{fig:model}b).  There
are several challenging aspects in this problem: (i) the TASEP of
particles of size $\ell$ (or $\ell$-TASEP) is known to have a
non-trivial current-density relation $j(\rho$), different from the
model with monomers
\cite{chou-lakatos:04,macdonald-gibbs-pipkin:68,shaw-zia-lee:03}; (ii)
even the simple on-off kinetics of dimers exhibits non-trivial
dynamics, for example the stationary state is reached from an empty
system through a double step relaxation process
\cite{mcghee-hippel:74,frey-vilfan:02}. The coupling of an equilibrium
process with two intrinsic relaxation regimes (on-off kinetics) to a
genuine driven process (the $\ell$-TASEP) suggests interesting
dynamical phenomena likely to result in new phases and regimes.

In absence of exact solution, the main challenge has been to construct
a refined mean field theory, based on probability theory, and to prove
the consistency of the approximation for the two competing process
(TASEP and the on-off kinetics). We have solved the mean field
equation within the mesoscopic limit in the stationary state:
\begin{eqnarray}
  \label{eq:dimers1}
  -\partial_x\left[\frac{\rho(1-2\rho)}{1-\rho}\right]+\left[\Omega_A\frac{(1-2\rho)^2}{(1-\rho)}-\Omega_D\rho\right]=0\, .
\end{eqnarray}
The complete equation is formally similar to the one for pointlike
particles (Eq.\ref{eq:pff1}): the current term (first bit in square
brackets) must be balanced by the on-off term (second part).

Exploiting the analytical properties of the solution of the mean field
equation and the local continuity of the current, we constructed the
global density profile.  We have performed extensive stochastic
simulations and the agreement with the mean field solution is
excellent. As in the case of TASEP/LK of monomers a new phase
coexistence region appears for some parameters.  The main effect of
extended nature of dimers on the phase behavior of the system is
related to the breaking of the (\emph{particle-hole}) symmetry of the
model.  This does have quantitative but not qualitative consequences
on the density profile and on the phase diagram, which remains
topologically unchanged. The origin of the \emph{robustness} of the
picture found for monomers can be traced back to the form of the
stationary density profile which depends exclusively on the form of
the current-density relation and of the isotherm of the on-off
kinetics. In both the monomer and the dimer case the current-density
relation is concave and presents a single maximum, while the isotherm
is unique and constant: these two features are enough to determine the
topology of the phase diagram. This robustness suggests that the TASEP
dynamics washes out the interesting two-step relaxation dynamics that
characterizes the on-off kinetics of dimers: the non-trivial outcome
is that, in these systems, the diffusion (yet asymmetric) always
dominates the large time-scale relaxation.

\subsection{Defect}
\begin{figure}[htbp]
\centering

 \psfrag{a}{(a)}
\psfrag{b}{(b)}
\psfrag{c}{(c)} 
 \psfrag{i=1}{$1$} 
 \psfrag{i=N}{$N$} 
 \psfrag{i=k}{$k$}
 \psfrag{L}{${\cal L}$}
 \psfrag{tau=1}{$r=1$}
 \psfrag{q}{$q<1$}
 \psfrag{omega_A}{$\omega_A$}
 \psfrag{omega_D}{$\omega_D$}
 \psfrag{alpha}{$\alpha$} \psfrag{beta}{$\beta$}
 \psfrag{alphaeff}{$\alpha_{\textrm{eff}}$}
 \psfrag{betaeff}{$\beta_{\textrm{eff}}$}
 \psfrag{ralpha}[l]{$\rho_1=\alpha$}
 \psfrag{rbeta}[d]{$\rho_N\!=\!1\!-\!\beta$}
 \psfrag{ralpha1}[l]{$\rho(0)=\alpha$}
 \psfrag{rbeta1}[d]{$\rho(1)\!=\!1\!-\!\beta$} \psfrag{q1}[l]{$\rho_k$}
 \psfrag{q2}[l]{$\rho_{k+1}$} \psfrag{q1a}[c]{$\frac{1}{1-q}$}
 \psfrag{q2a}[c]{$\frac{q}{1-q}$} \psfrag{q}{$q$} \psfrag{x=k}{$i=k$}
 \psfrag{I}{L} \psfrag{II}{R}
 \psfrag{x}{$x$}
 \psfrag{j}{$j$}
 \psfrag{r}{$\rho$}
 
\includegraphics[width=\textwidth]{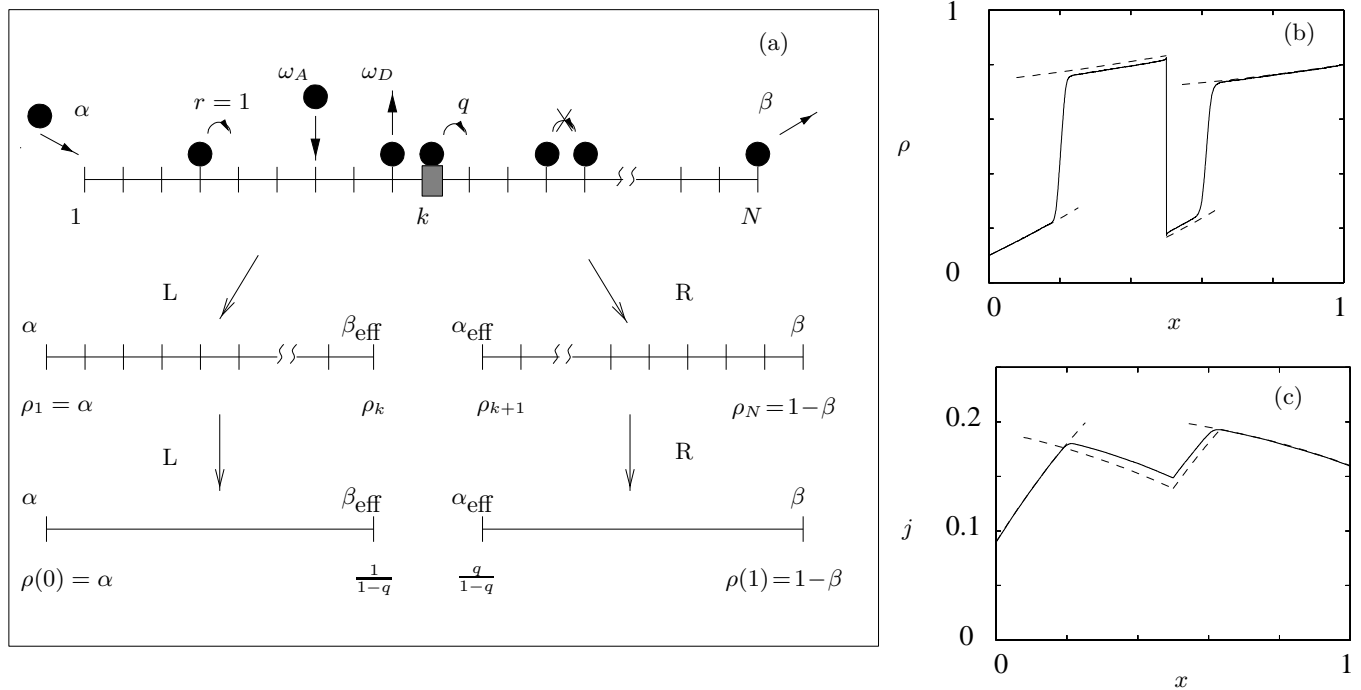}
\caption{(a) Cartoon representing the TASEP/LK with a bottleneck: 
above a threshold density the system can be divided into two lattices
where the results of TASEP/LK are valid. (b) Density and (c) current
in a case where the bottleneck is relevant and competes with the
boundaries leading to the formation of a double domain wall,
theoretical results (dashed lines) are superimposed to numerical
simulations (bold line).}
\label{fig:defect}       
\end{figure}

Another question that often arises in the study of these models
concerns the role of some kind of randomness: the motion of the
particles can be altered by structural defects of the track or by
microtubule associated proteins (MAP). A good modeling requires to
introduce some sort of disorder in the system: site-related or
particle related, quenched or annealed. While disorder in TASEP has
been treated with exact and approximated methods
\cite{harris-stinchcombe:04,
  juhasz-santen-igloi:05,kolwankar-punnoose:00},the role of
bottlenecks was never coupled to the TASEP/LK. As a preliminary study
of the role of quenched disorder, it becomes particularly interesting
to investigate the influence of an isolated defect (i.e.\ point-wise
disorder) on the stationary properties of the TASEP/LK. In
Ref.\cite{pierobon-etal:06} we extensively studied this model where
the defect has been characterized by a reduced hopping rate $q<1$ (see
Fig.~\ref{fig:defect}).

As a consequence of the competition between the TASEP and LK dynamics,
the effects of a single bottleneck in the TASEP/LK model are much more
dramatic than in the simple TASEP \cite{kolomeisky:98} (or the TASEP
for extended objects \cite{shaw-kolomeisky-lee:03}), where a localized
defect was shown to merely shift some transition lines in the
phase-diagram, but do not affect its topology. Here, new and mixed
phases induced by the bottleneck have been obtained.

As a key concept of our analysis, we have introduced the {\it carrying
  capacity}, which is defined as the maximal current that can flow
through the bulk of the system. In contrast to the simple TASEP the
spatial dependence of the current, caused by the Langmuir kinetics,
makes the carrying capacity non-trivial: the defect depletes the
current profile within a distance that we called \emph{screening
  length}. This quantity increases with the strength of the defect and
decreases with the attachment-detachment rates. The competition
between the current imposed at the boundaries and the one limited by
the defect determines the density profiles and the ensuing
phase-diagram.  When the boundary currents are dominant, the phase
behavior of the defect-free system is recovered. Also, above some
critical entrance and exit rates, the system transports the maximal
current, independently of the boundaries.  Between these two extreme
situations, we have found several coexistence phases, where the
density profile exhibits stable shocks and kinks. Indeed, above some
specific parameter values the phase-diagram is characterized by
\emph{bottleneck phases}. Depending on the screening length imposed by
the defect, which can cover the entire system or part of it, different
phase-diagrams arise. The latter are characterized by four, six or
even nine bottleneck phases, which have been quantitatively studied
within the mean-field theory introduced in Sec.\ref{sec:pff}: in fact
for entrance and exit rates that exceed the critical value $q/(1+q)$
(for which the bottleneck becomes relevant) the system can be split in
two sublattices that can be treated as independent TASEP/LK with
effective exit/entrance rates at the junction
(Fig.~\ref{fig:defect}a).

Our results were checked against numerical simulations, which brings
further arguments in favor of the validity of mean-field approaches
for studying the TASEP/LK-like models (Figs.\ref{fig:defect}b-c).  The
somewhat surprising quantitative validity of this approximate scheme
can be traced back to the current-density relationship, which is
correctly predicted by the mean-field theory.

Eventually, we think that this study showed clearly that the presence
of disorder in the TASEP/LK model, even in its simplest form,
generally gives rise to quite rich and intriguing features and should
motivate further studies of more `realistic' and biophysically
relevant situations, as in the presence of clusters of competing
defects or quenched site-wise randomness (where the motors are slowed
down at several points in the system).

\section{Towards the experiments}
\subsection{Tracer dynamics}
All the models described so far would be rather useless without the
possibility to measure quantitatively the density and to localize the
shock. To this aim, simple bulk fluorescence imaging would be
difficult to apply to quantitative experiments: single molecule
analysis are far more promising. We proposed hence a method, based on
single particle tracking, to measure the density of the system
\cite{phd:pierobon}. The idea is to use the simple TASEP (exact)
results to reconstruct the density from the velocity or the diffusion
constant of the tracer. It is known that the average velocity
$v\equiv\frac{d\avg{x(t)}}{dt}$ is related to the density $\rho$
through the relation $\rho=1-v$. Exact results on a ring show that the
same relation holds for the diffusion constant
$D\equiv\frac{\avg{x^2}-\avg{x}^2}{t}=1-\rho$
\cite{derrida-evans-mukamel:93}.

\begin{figure}[htbp]
\centering
\includegraphics[width=\textwidth]{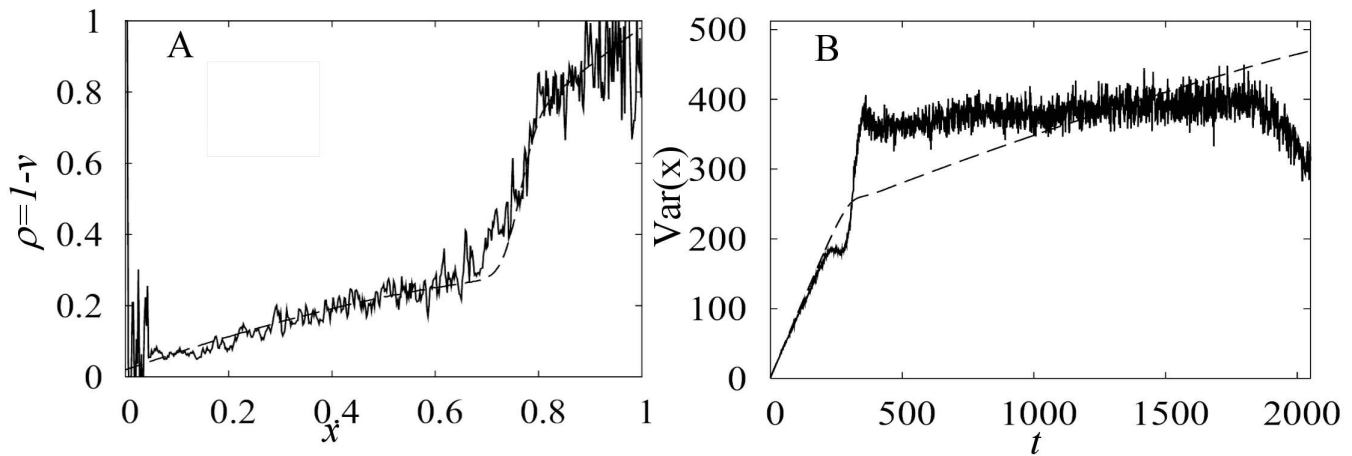}
\caption{(a) Density profile of TASEP/LK derived from the analysis of
  100 tracer particles: wiggly line match rather well the theoretical
  prediction (dashed line). (b) Superposition of the first two moments
  of $P(x,t)$, $\avg{x}(t)$ (dashed) and $Var[x(t)]$ (wiggly) in time:
  according to the simple TASEP results the two quantities should be
  the same; this is true only until the particles have reached the
  shock, afterwards the variance deviates.}
\label{fig:tracer}      
\end{figure}

Since the rates scales with the size of the system, according to the
mesoscopic limit, we suppose that the influence of attachment
detachment on this relation to be negligible and the density to be
locally continuous. We simulated the system TASEP/LK and measured the
position $x$ as a function of the time for several particles, in order
to construct the probability density function $P(x,t,|0,0)$ to find a
tracer particle at site $x$ after a time $t$ from its entrance in the
system at site $0$.  From this function we can measure the velocity
and the diffusion constant. As shown in Fig.~\ref{fig:tracer}, the
shock is localized within a $10\%$ precision through both the
indicators. While the density is well reconstructed from the
information on the velocity (Fig.~\ref{fig:tracer}a), the hypothesis
$\rho=1-D$ works well only in the low density phase before the
particle arrives at the shock. Once the particle has passed through
the shock the hypothesis on the continuity of the density breaks down
and the relation is not valid anymore while the loss of particles in
the high density phase makes the system subdiffusive. Yet, the
quantity $1-D$ can be used to localized the shock
(Fig.~\ref{fig:tracer}b).

Numerical simulations show that, on system of realistic size (i.e.
$1\mu m$, roughly 100 sites), analysing the trajectories of a hundred
particles is enough to reconstruct the density profile and localize
the shock (after a time moving average resulting on a smoothing of the
data).  This suggests that single molecule experiments observing local
features could give information on the global scale.

\subsection{Single molecule in the cell}
So far much information concerning processive molecular motors was
provided by single molecule experiments. Typically a latex bead is
attached to the motor to make it visible and to manipulate it: this
allowed us to measure not only velocity and processivity of the motors,
but even the forces they exert, the steps and (in some lucky cases)
substeps \cite{veigel-etal:02, uemura-etal:04, dunn-spudich:07,
  cappello-etal:07}.  In-vitro observations can be quite
controversial: some experiments did not show any difference upon
increasing the concentrations of motors, as if traffic effects were
not relevant \cite{seitz-surrey:06}.  However the situation in the
overcrowded environment of the cell is hardly reproducible in vitro
and could reserve more surprises.

\begin{figure}[htbp]
\centering
\includegraphics[width=\textwidth]{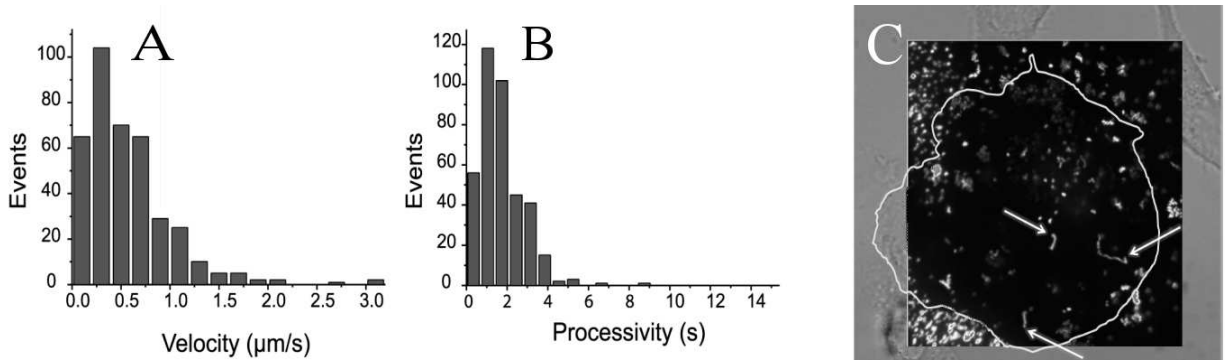}
\caption{Histograms of the speed (A) and processivity (B) of kinesin
  in Hela cells. The results confirmed in vitro measurements.  (C)
  Picture of the trajectories (arrows) of a QD labeled Kinesins: image
  obtained choosing the maximal intensity over many frames of a movie
  with long integration time (100ms).}
\label{fig:histo}       
\end{figure}

Since single molecule experiments have often been criticized because
of their distance from the real systems, many groups are cautiously
moving the single molecule experiments directly into the cell
\cite{dahan-etal:03,selvin-etal:05,xie-yu-yang:06,watanabe-higuchi:07}.
In this situation the main problem is to have a good signal-noise
ration which is hardly achievable with the usual fluorescent probes.
Recently a method to mark and observe single molecule in vivo by using
quantum dots (QDs) has been proposed
\cite{courty-etal:06,xie-yu-yang:06}. In contrast with the usual
fluorescent probes, the QDs do not bleach, are excitable on all the
visible spectrum and show a very narrow emission band. The only
drawback is that they blink without a typical timescale: this
inconvenient can be avoid by taking longer series of images. In
Ref.\cite{courty-etal:06} kinesins are biotinilated to bind to a
streptavidinated QD. The conjugated particles so obtained are
introduced in Hela cell by \emph{pinocytosis} followed by osmotic
shock and the motion of the QDs is observed with a customized
fluorescence microscopy setup and a fast CCD camera. The QDs signal on
25 pixels is fit with a Gaussian to obtain sub-pixel resolution (FIONA
\cite{yildiz-etal:03}). The blinking of moving QDs is taken as an
evidence that we are working in single molecule. The results on
kinesin speed ($570\pm20nm/s$) and processivity ($1.73\pm0.06s$,
i.e.$\sim1\mu m$ in space) are compatible to the ones known from
in-vitro experiments (Fig.~\ref{fig:histo}a-b) \footnote{Additional
  experiments, using drugs, were carried out to ensure that the motion
  was actually due to motors.}.

\section{Conclusion}

In this proceeding we have reviewed a model introduced to describe
intracellular transport \cite{parmeggiani-franosch-frey:03} and its
possible extensions towards a more realistic picture: we investigated
how the introduction of dimers \cite{pierobon-frey-franosch:06} and
the presence of disorder on the track \cite{pierobon-etal:06} would
affect the known results.  These theoretical approaches while aiming
to a realistic description of the traffic phenomena, are inspiring
statistical models interesting in its own right (see e.g.
\cite{chowdhury-schadschneider-nishinari:05,mobilia-etal:06}).  The
latest experimental successes in tracking individual motors in living
cells combined with an appropriate analysis,
presented in the last section, could not only confirm the theories but
also provide ispiration for a complete picture of the cell logistic.

\subparagraph{Acknowledgments} I would like to thank the coauthors of
Refs.\cite{pierobon-frey-franosch:06,pierobon-etal:06,cappello-etal:07}.
I benefit from useful discussions with A. Parmeggiani, S. Achouri, A.
Dunn and L.  Sengmanivong.

%
%
%



\end{document}